\RequirePackage{lineno}
\documentclass[aps,prb,twocolumn,amssymb,amsmath,superscriptaddress,floatfix]{revtex4}
\usepackage{graphicx}
\usepackage{bm}
\usepackage{multirow}
\usepackage{appendix}
\usepackage{color}

\begin{document}


\title {The possible frustrated superconductivity in the kagome superconductors}

\author{Hong-Min Jiang}
\email{monsoonjhm@sina.com} \affiliation{School of Science, Zhejiang
University of Science and Technology, Hangzhou 310023, China}
\affiliation{HK Institute of Quantum Science $\&$ Technology and
Department of Physics, The University of Hong Kong, Pokfulam Road,
Hong Kong, China}
\author{Wen-Qian Dong}
\affiliation{School of Science, Zhejiang University of Science and
Technology, Hangzhou 310023, China}
\author{Shun-Li Yu}
\email{slyu@nju.edu.cn} \affiliation{National Laboratory of Solid
State Microstructures and Department of Physics, Nanjing University,
Nanjing 210093, China} \affiliation{Collaborative Innovation Center
of Advanced Microstructures, Nanjing University, Nanjing 210093,
China}
\author{Z. D. Wang}
\email{zwang@hku.hk} \affiliation{HK Institute of Quantum Science
$\&$ Technology and Department of Physics, The University of Hong
Kong, Pokfulam Road, Hong Kong, China} \affiliation{Hong Kong Branch
for Quantum Science Center of Guangdong-Hong Kong-Macau GBA, China}

\date{\today}

\begin{abstract}
Geometric frustration has long been a subject of enduring interest
in condensed matter physics. While geometric frustration
traditionally focuses on magnetic systems, little attention is paid
to the "frustrated superconductivity" which could arise when the
superconducting interaction conflicts with the crystal symmetry. The
recently discovered kagome superconductors provide a particular
opportunity for studying this due to the fact that the frustrated
lattice structure and the interference effect between the three
sublattices can facilitate the frustrated superconducting
interaction. Here, we propose a theory that supports the frustrated
superconducting state, derived from the on-site $s$-wave
superconducting pairing in conjunction with the nearest-neighbor
pairings hopping and the unique geometrical frustrated lattice
structure. In this state, whereas the mutual $2\pi/3$ difference of
the superconducting pairing phase causes the six-fold modulation of
the amplitude and breaks the time-reversal symmetry with $4\pi$
phase changes of the superconducting pairing as one following it
around the Fermi surface, it is immune to the impurities without the
impurity-induced in-gap states and produces the pronounced
Hebel-Slichter peak of the nuclear spin-lattice relaxation rate
below $T_{c}$. Notably, the theory also reveals a disorder-induced
superconducting pairing transition from the frustrated
superconducting state to an isotropic $s$-wave superconducting state
without traversing the nodal points, recovering and explaining the
behavior found in experiment. This study not only serves as a
promising proposal to mediate the divergent or seemingly
contradictory experimental outcomes regarding superconducting
pairing symmetry, but may also pave the way for advancing
investigations into the frustrated superconducting state.

\end{abstract}


\maketitle

\section{Introduction}
The discovery of superconductivity in a family of compounds
AV$_{3}$Sb$_{5}$ (A=K, Rb, Cs), which share a common lattice
structure with kagome net of vanadium atoms, has set off a new boom
of researches on the superconductivity as well as the associated
exotic quantum states
~\cite{Ortiz1,SYYang1,Ortiz2,QYin1,KYChen1,YWang1,ZZhang1,YXJiang1,
FHYu1,XChen1,HZhao1,HChen1,HSXu1,Liang1,CMu1,CCZhao1,SNi1,WDuan1,
Xiang1,PhysRevX.11.041030,PhysRevB.104.L041101,
ZLiu1,NatPhys.M.Kang,YFu1,HTan1,Shumiya1,FHYu2,LYin1,Nakayama2,
KJiang1,LNie1,HLuo1,Neupert1,YSong1,Nakayama1,HLi1,CGuo1,HLi2,HLi3,XWu1,Denner1,SCho1,YPLin3,
LZheng1,CWen1,Tazai1,Jiang1,Mielke1}. The appealing aspects of these
compounds lie in that they incorporate many remarkable properties of
the electron structure, such as VHF, FS nesting and nontrivial band
topology~\cite{Ortiz1}. Meanwhile, the superconducting (SC) phase
emerges next to a charge density wave phase in the
pressure-temperature phase diagram. As the electrons in these
materials suffer simultaneously from the geometrical frustration,
topological band structure and the competition between different
possible ground states, the observations of the superconductivity in
these topological metals are in themselves exotic and rare. The
connection to the underlying lattice geometry and the topological
nature of the band structure further place them in the context of
wider research efforts in topological physics and superconductivity.

To understand the underlying mechanism of the superconductivity in
this family of kagome superconductors, the determination of the
pairing symmetry of the SC order parameter is thought to be a
prerequisite step. For this goal, numerous experiments with various
means have been conducted in the past several years. However,
inconsistent results were found so far in experimental measurements
and data analysis, and some unusual features were discovered as well
in the SC state of AV$_{3}$Sb$_{5}$. On the one hand, the SC
pairings in these compounds are suggested to be of the conventional
$s$-wave type, supported by the appearance of the Hebel-Slichter
coherence peak just below $T_{c}$ in the nuclear magnetic resonance
spectroscopy~\cite{CMu1} and the nodeless SC gap in both the
penetration depth measurements~\cite{WDuan1} and the angle-resolved
photoemission spectroscopy (ARPES) experiment~\cite{YZhong1}.
Intriguingly, recent magnetic penetration depth measurements also
revealed a impurity induced transition from an anisotropic
fully-gapped state to an isotropic full-gap state without passing
through a nodal state~\cite{Roppongi1}. On the other hand, the
indications of time-reversal symmetry breaking discovered in the SC
state~\cite{Mielke1,YWang1,YWu1}, together with the nodal SC gap
feature detected by some
experiments~\cite{HZhao1,CCZhao1,HSXu1,Liang1,HChen1}, hint to an
unconventional superconductivity.

Although the majority of experimental evidence in the
AV$_{3}$Sb$_{5}$ materials indicates that the superconductivity has
a conventional origin with spin singlet $s$-wave
pairing~\cite{CMu1,WDuan1,YZhong1,YXie1}, the field is haunted by
the seemingly divergent experimental observations regarding the
pairing symmetry. Theoretically, a variety of possible pairing
symmetries have been put forward in the kagome systems, ranging from
the $s$-wave~\cite{WSWang1,THan1,Tazai1}, $d$-wave, and extended
$s_{ext}$-wave to $d_{x^{2}-y^{2}}+id_{xy}$-wave pairing
symmetries~\cite{SLYu1,WSWang1,Romer1,Kiesel3}, and even triplet
$p$-wave~\cite{Tazai1} and $f$-wave pairing
symmetries~\cite{SLYu1,CWen1,XWu1,TZhou1}. Considering the interplay
of the superconductivity and CDW phase, theoretical analysis has
also shown that a conventional fully gapped superconductivity is
unable to open a gap on the domains of the chiral $2\times 2$ charge
density wave (CDW) and results in the gapless edge modes in the SC
state~\cite{YGu1}. A more direct consideration of the interplay
between the superconductivity and the chiral $2\times 2$ CDW has
revealed that a modulated and even nodal SC gap features show up
albeit an on-site $s$-wave SC order parameter being included in the
study~\cite{HMJiang2,HMJiang3,TZhou2}. In this scenario, the
coexistence of the conventional superconductivity and the chiral CDW
has also been proposed to explain time-reversal symmetry breaking in
the kagome superconductors~\cite{HMJiang2,HMJiang3,TZhou1}.

Nevertheless, more recent experiments have evidenced that the
time-reversal symmetry breaking has a direct connection to the SC
state~\cite{YWang1,TLe1}, especially in the case where the CDW order
was fully suppressed by Ta atoms doping into
CsV$_{3}$Sb$_{5}$~\cite{HDeng1}. While a variety of possible pairing
states with time-reversal symmetry breaking have been proposed,
there is still limited knowledge as to how the SC pairings that
possess the conventional origin could have the intrinsic
time-reversal symmetry breaking SC state.

In this paper, we propose a theory in which the conventional on-site
$s$-wave SC pairing transforms to the time-reversal symmetry
breaking SC state, as a result of the nearest-neighbor pairings
hoping and the unique geometrical frustrated lattice structure. In
this state, the phase of the SC pairing on the three sublattices
mutually differs by $2\pi/3$, an analogy to the frustrated
magnetism, which is dubbed as "frustrated SC state" thereafter. The
amplitude of the frustrated SC state modulates along the Fermi
surface six-fold symmetrically without passing through the nodes.
The frustrated SC state breaks the time-reversal symmetry explicitly
with the $4\pi$ phase changes of the SC pairing as one following it
around the Fermi surface, i.e., winding number of $2$. Though the
frustrated SC state has the structure along the Fermi surface
exactly as the $d_{x^{2}-y^{2}}+id_{xy}$-wave pairing, it is immune
from the impurities without the impurity-induced in-gap states and
produces the pronounced typical Hebel-Slichter peak of the nuclear
spin-lattice relaxation rate below $T_{c}$. Particularly, the theory
also reveals a disorder-induced SC pairing transition from the
frustrated SC state to an isotropic $s$-wave SC state without
passing through the nodal point, recovering the behavior found in
experiment. This study not only serves as a promising proposal to
mediate the divergent or seemingly contradictory experimental
outcomes regarding SC pairing symmetry, but may also paves the way
for further studies on the frustrated SC state.

The remainder of the paper is organized as follows. In Sec. II, we
introduce the model Hamiltonian and carry out analytical
calculations. In Sec. III, we present numerical calculations and
discuss the results. In Sec. IV, we make a conclusion.

\section{model and method}
Considering the SC pairing hoppings among the nearest-neighbor
sites, the Hamiltonian of the system is modified by adding a pairing
hopping term,
\begin{eqnarray}
H_{PH}&=&J\sum_{\langle
ij\rangle}(c^{\dag}_{i\uparrow}c^{\dag}_{i\downarrow}
c_{j\downarrow}c_{j\uparrow}+h.c.)
\end{eqnarray}
where $c^{\dag}_{i\sigma}$ creates an electron with spin $\sigma$ on
the site $i$ of the kagome lattice and $\langle ij\rangle$ denotes
nearest-neighbors (NN). Mimicking the NN magnetic interactions, we
may define the SC operators $\tilde{\Delta}^{\dag}_{i}\equiv
c^{\dag}_{i\downarrow}c^{\dag}_{i\uparrow}$ and
$\tilde{\Delta}_{j}\equiv c_{j\uparrow}c_{j\downarrow}$. Then, the
pairing hopping term can be rewritten as,
\begin{eqnarray}
H_{PH}&=&J\sum_{\langle
ij\rangle}(\tilde{\Delta}^{\dag}_{i}\tilde{\Delta}_{j}+H.c.).
\end{eqnarray}
On the geometry frustrated kagome lattice with positive pairing
hopping $J$, the pairings on the three sublattices with the mutual
phase differences $2\pi/3$ result in a maximal energy gain
$-3J|\Delta|^{2}$, where we assume the pairing amplitudes on the
three sublattice sites being the same. By comparison, while the
energy gain is $-2J|\Delta|^{2}$ for the pairings on the three
sublattices with one of them possessing the opposite phase, there is
energy lost $6J|\Delta|^{2}$ for the pairings with identical phase.
Thus, we are encountered with the SC version of the spin
frustration, i.e., the frustrated superconductivity.

To exemplify the frustrated superconductivity, we put the above
argument into a kagome model Hamiltonian that is generally used in
the previously investigations on the kagome superconductors. The
total Hamiltonian reads,
\begin{eqnarray}
H&=&H_{0}+H_{P}+H_{PH}.
\end{eqnarray}
The first term of $H$ is the single orbital tight-binding
Hamiltonian involving the effective electron hoppings on a kagome
lattice:
\begin{eqnarray}
H_{0}&=&-\sum_{\langle
ij\rangle\sigma}(t_{ij}c^{\dag}_{i\sigma}c_{j\sigma}+H.c.)
-\mu\sum_{i\sigma}c^{\dag}_{i\sigma}c_{i\sigma},
\end{eqnarray}
where $t_{ij}$ is the hopping integral between the NN sites, and
$\mu$ is the chemical potential.

The second part of $H$ describes the effective SC pairing. It is
expressed as,
\begin{eqnarray}
H_{P}&=&\sum_{\langle ij\rangle}(\Delta_{i}
c^{\dag}_{i\uparrow}c^{\dag}_{i\downarrow}+H.c.).
\end{eqnarray}
Here, we choose the on-site $s$-wave SC order parameter
$\Delta_{i}=V_{s}\langle c_{i\uparrow}c_{i\downarrow}\rangle$.

In the mean-field approximation, the Hamiltonian $H$ can be written
as,
\begin{eqnarray}
H=&&-\sum_{\langle
ij\rangle\sigma}(t_{ij}c^{\dag}_{i\sigma}c_{j\sigma}+H.c.)
-\mu\sum_{i\sigma}c^{\dag}_{i\sigma}c_{i\sigma} \nonumber\\
&&+\sum_{i}[(\Delta_{i}-\frac{J}{V_{s}}\eta_{i})
c^{\dag}_{i\uparrow}c^{\dag}_{i\downarrow}+H.c.] \nonumber\\
&&+\frac{1}{V_{s}}\sum_{i}\Delta^{\ast}_{i}\Delta_{i}-\frac{J}{V_{s}^{2}}\sum_{\langle
ij\rangle}(\Delta^{\ast}_{i}\Delta_{j}+H.c.),
\end{eqnarray}
where $\eta_{i}=\sum_{\delta}\Delta_{\delta}$ is defined as the
summation of the NN site pairing order parameters around site $i$.

For the translational invariant system, the total Hamiltonian $H$
can be written in the momentum space as,
\begin{eqnarray}
H(\mathbf{k})&=&\sum_{\mathbf{k}}\hat{\Psi}^{\dag}_{\mathbf{k}}\hat{\mathcal{H}}_{\mathbf{k}}
\hat{\Psi}_{\mathbf{k}},
\end{eqnarray}
with
$\hat{\Psi}_{\mathbf{k}}=(c_{A,\mathbf{k}\uparrow},c_{B,\mathbf{k}\uparrow},c_{C,\mathbf{k}\uparrow},
c^{\dag}_{A,\mathbf{-k}\downarrow},c^{\dag}_{B,\mathbf{-k}\downarrow},c^{\dag}_{C,\mathbf{-k}\downarrow})^{T}$
and
\begin{eqnarray}
\hat{\mathcal{H}}_{\mathbf{k}}=\left(
\begin{array}{cc}
\hat{\mathcal{H}}^{0}_{\mathbf{k}} & \hat{\Delta} \\
\hat{\Delta}^{\ast} & -\hat{\mathcal{H}}^{0}_{\mathbf{k}}
\end{array}
\right),
\end{eqnarray}
where
\begin{eqnarray}
\hat{\mathcal{H}}^{0}_{\mathbf{k}}=\left(
\begin{array}{ccc}
-\mu & -2t\cos k_{1} &
-2t\cos k_{2} \\
-2t\cos k_{1} & -\mu & -2t\cos k_{3} \\
-2t\cos k_{2} & -2t\cos k_{3} & -\mu
\end{array}
\right),
\end{eqnarray}
and
\begin{eqnarray}
\hat{\Delta}=\left(
\begin{array}{ccc}
\Delta_{A}-\frac{J}{V_{s}}\eta_{A} & 0 & 0 \\
0 & \Delta_{B}-\frac{J}{V_{s}}\eta_{B} & 0 \\
0 & 0 & \Delta_{C}-\frac{J}{V_{s}}\eta_{C}
\end{array}
\right).
\end{eqnarray}
Here the index $l=A,B,C$ in $c_{lk\sigma}$ labels the three basis
sites in the triangular primitive unit cell as shown in
Fig.~\ref{fig1}(a). $k_{n}$ is abbreviated from
$\mathbf{k}\cdot\mathbf{\tau}_{n}$ with
$\mathbf{\tau}_{1}=\hat{x}/2$,
$\mathbf{\tau}_{2}=(\hat{x}+\sqrt{3}\hat{y})/4$ and
$\mathbf{\tau}_{3}=\mathbf{\tau}_{2}-\mathbf{\tau}_{1}$ denoting the
three NN vectors. $\Delta_{\alpha}=V_{s}\langle
c_{\alpha,k\uparrow}c_{\alpha,-k\downarrow}\rangle$ and
$\eta_{\alpha}=\sum_{\delta}\Delta_{\alpha+\delta}$. In the
procedure for obtaining Eq. (7), we also neglect the constant terms
$\frac{1}{V_{s}}\sum_{i}\Delta^{\ast}_{i}\Delta_{i}$ and
$\frac{J}{V_{s}^{2}}\sum_{\langle
ij\rangle}(\Delta^{\ast}_{i}\Delta_{j}+H.c.)$.

The Hamiltonian $H_{\mathbf{k}}$ can be solved by resorting the
Bogoliubov transformation which leads to the following Bogoliubov-de
Gennes equations,
\begin{eqnarray}
\sum_{\mathbf{k}}\left(
\begin{array}{lr}
\hat{\mathcal{H}}^{0}_{\mathbf{k}} & \hat{\Delta} \\
\hat{\Delta}^{\ast} & -\hat{\mathcal{H}}^{0}_{\mathbf{k}}
\end{array}
\right)\left(
\begin{array}{lr}
u_{n,\mathbf{k},\sigma} \\
v_{n,\mathbf{k},\bar{\sigma}}
\end{array}
\right)= E_{n,\mathbf{k}}\left(
\begin{array}{lr}
u_{n,\mathbf{k},\sigma} \\
v_{n,\mathbf{k},\bar{\sigma}}
\end{array}
\right),
\end{eqnarray}
where $u_{n,\mathbf{k},\sigma}$ and $v_{n,\mathbf{k},\bar{\sigma}}$
are the Bogoliubov quasiparticle amplitudes with momentum
$\mathbf{k}$ and eigenvalue $E_{n,\mathbf{k}}$. The amplitudes of
the SC pairing and the electron densities are obtained through the
following self-consistent equations,
\begin{eqnarray}
\Delta&=&\frac{V_{s}}{2N}\sum_{n,\mathbf{k}}u_{n,\mathbf{k},\sigma}
v^{\ast}_{n,\mathbf{k},\bar{\sigma}}
\tanh(\frac{E_{n,\mathbf{k}}}{2k_{B}T}) \nonumber\\
n_{e}&=&\frac{1}{N}\sum_{n,\mathbf{k}}\{|u_{n,\mathbf{k},\uparrow}|^{2}
f(E_{n,\mathbf{k}})+|v_{n,\mathbf{k},\downarrow}|^{2}[1-f(E_{n,\mathbf{k}})]\},
\end{eqnarray}
with $N$ denoting the total meshes in the momentum space.

\section{Results and discussion}

\subsection{The self-consistent results of the SC pairing}

In the calculations, we choose the hopping parameter $t=1$ as the
energy unit, and fixes temperature $T=1\times10^{-5}$, unless
otherwise specified. The chemical potential $\mu$ is adjusted to
maintain a band filling of 1/6 hole doping, corresponding to the van
Hove singularity filling. The SC order parameter is determined
self-consistently by treating the pairing hopping strength $J$ as a
variational parameter, while fixing the pairing interaction at
$V_{s}=1.4$. In cases where multiple solutions arise from the
self-consistent calculations at the same temperature but with
different sets of initially random input parameters, we compare
their free energy defined as
\begin{eqnarray}
F=&&-2k_{B}T\sum_{\mathbf{k},E^{\mathbf{k}}_{n}>0}\ln[2\cosh(\frac{E^{\mathbf{k}}_{n}}{2k_{B}T})]
+N\frac{|\Delta_{i}|^{2}}{V_{s}} \nonumber\\
&&-\frac{J}{2V_{s}^{2}}\sum_{\mathbf{k},\langle
ij\rangle}(\Delta^{\ast}_{i}\Delta_{j}+H.c.),
\end{eqnarray}
so as to find the most favorable state in energy.

\vspace*{.0cm}
\begin{figure}[htb]
\begin{center}
\vspace{-.2cm}
\includegraphics[width=240pt,height=90pt]{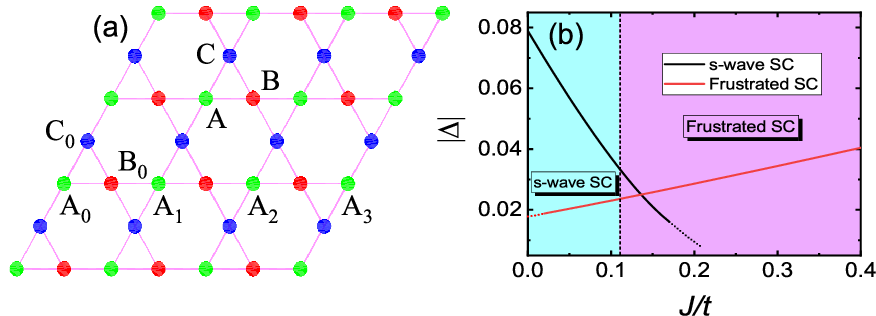}
\caption{(a) Structure of the kagome lattice, made out of three
sublattices A (green dots), B (red dots), and C (blue dots). (b) The
SC pairing amplitude as a function of the pairing hopping strength
$J$. The light cyan region with $J<0.11$ indicates the conventional
$s$-wave SC state, and the light purple region with $J>0.11$ denotes
the frustrated SC state.}\label{fig1}
\end{center}
\end{figure}

As a function of the pairing hopping strength $J$, the calculations
show that the conventional $s$-wave SC state possesses the lower
free energy when $J<0.11$, whereas the frustrated SC state wins over
when $J>0.11$. In the conventional $s$-wave SC regime, the pairing
hopping tends to suppress the SC pairing amplitude, but in the
frustrated SC region the pairing hopping enhances the SC pairing
amplitude. The results are displayed in Fig.~\ref{fig1}(b), where
the vertical dashed line borders the transition between the
conventional $s$-wave and frustrated SC states, and the dotted
portions in each curves are numerically unreachable in their
respective state.

\subsection{The structure of the frustrated SC pairing}
The SC pairings in the Hamiltonian are defined on the orbital basis,
while the spectral experiments usually measure the electronic
structures near the Fermi surface, which is certainly on the band
basis. To gain the clear picture about the frustrated SC state, we
transform the SC pairings into band space through a unitary
transformation,
\begin{eqnarray}
\Delta_{mn}(k)&=&\sum_{\alpha\beta}u^{\ast}_{m\alpha}(k)\Delta_{\alpha\beta}u^{\ast}_{n\beta}(-k),
\end{eqnarray}
where $u_{m\alpha}(k)$ and $\Delta_{\alpha\beta}$ are the
eigenstates of band $m$ and the SC pairings defined on the orbital
basis. The diagonal components of $\Delta_{mn}(k)$ correspond to the
intra-band pairing, and the off-diagonal components represent the
inter-band pairings. While both the diagonal and the off-diagonal
components could be non-zero, the inter-band pairings in this
circumstance do not lead to a gap opening at the Fermi level.
Meanwhile, the first band $(m=1)$ and third band $(m=3)$ stay away
from the Fermi level. Thus, only the diagonal component in the
middle band $\Delta_{22}(k)$ is relevant to the experimental
measurements.

\vspace*{.0cm}
\begin{figure}[htb]
\begin{center}
\vspace{-.2cm}
\includegraphics[width=240pt,height=180pt]{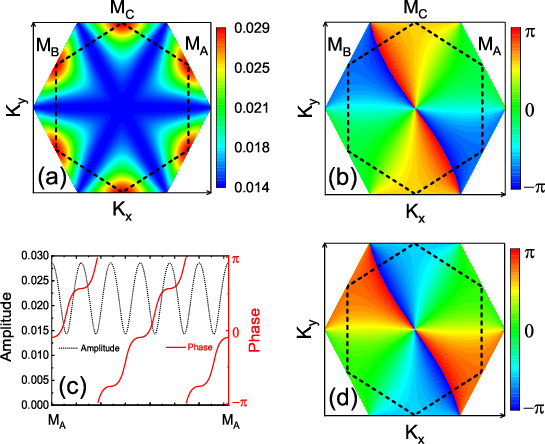}
\caption{Distributions of the amplitude (a) and phase (b) of the
frustrated SC state in the primitive Brillouin zone. (c) Variations
of the amplitude and phase of the frustrated SC state following in a
counterclockwise direction around the Fermi surface shown as dashed
lines in (a) and (b). (d) Same as (b) but with opposite chirality of
the SC pairing.}\label{fig2}
\end{center}
\end{figure}

Figs.~\ref{fig2}(a) and (b) present respectively the distributions
of the amplitude and the phase of $\Delta_{22}(k)$ for the
frustrated SC state in primitive Brillouin zone. Fig.~\ref{fig2}(c)
displays their distributions following a counterclockwise direction
around the Fermi surface. Two salient features are observed from the
figure. Firstly, while the SC state is fully gaped, the amplitude of
the SC pairing modulates along the Fermi surface with six-fold
symmetry, which is in line with the recent experimental
surveys~\cite{Roppongi1}. Due the sublattice interference, the van
Hove singular points near the van Hove filling come exclusively from
the same sublattice sites, and so is the SC pairings. The same phase
of the SC pairings on the same sublattice sites gives rise to the
maximum $|\Delta|$ at the $M$ points. On the other hand, the
midpoint between the two adjacent van Hove points originates from
the equally mixing of two inequivalent sublattice sites. This means
that $\Delta_{22}(k)$ at the midpoint is derived from the equally
mixing of the SC pairings on two inequivalent sublattice sites,
i.e., $(|\Delta|+|\Delta|e^{\pm i\frac{2\pi}{3}})/2$ or
$(|\Delta|e^{i\frac{2\pi}{3}}+|\Delta|e^{-i\frac{2\pi}{3}})/2$,
which produce exactly $|\Delta|/2$ for the minimum of
$|\Delta_{22}(k)|$ at the midpoint, just as depicted in
Figs.~\ref{fig2}(a) and (c). Secondly, depending on the chirality,
the phase of the SC pairing undergoes $4\pi$ [Fig.~\ref{fig2}(b)] or
$-4\pi$ [Fig.~\ref{fig2}(d)] changes, i.e., winding of $\pm2$, when
one follow it in a counterclockwise direction around the Fermi
surface, explicitly demonstrating the breaking of the time reversal
symmetry.

\subsection{Conventional aspects of the frustrated SC state}
\begin{figure}[htb]
\begin{center}
\vspace{-.2cm}
\includegraphics[width=240pt,height=170pt]{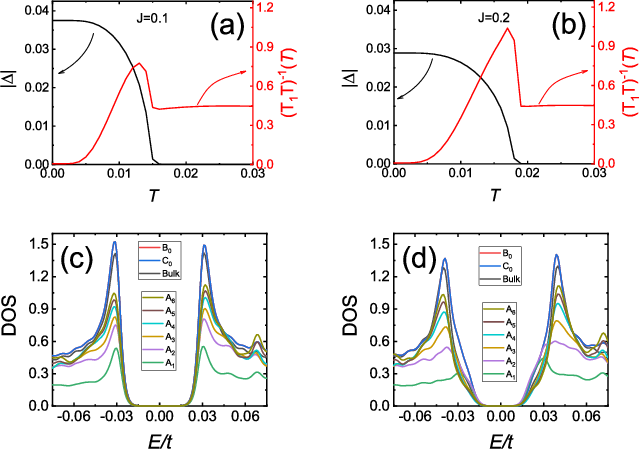}
\caption{Temperature dependence of the amplitude of SC pairing
$|\Delta|$ and $(T_{1}T)^{-1}$ for $J=0.1$ (a) and $J=0.2$ (b). The
energy dependence of the DOS on a series of sites for $J=0.1$ (c),
and $J=0.2$ (d). In (c) and (d), $B_{0}$ and $C_{0}$ denote the
other two sublattice sites within the same unit cell as the impurity
site $A_{0}$, and $A_{1}-A_{5}$ stand for the LDOS at the same
sublattice sites as the impurity site $A_{0}$ in different unit cell
moving away from $A_{0}$ [See Fig.~\ref{fig1}(a).].}\label{fig3}
\end{center}
\end{figure}

The most notable characteristics in supporting the conventional SC
pairing for the kagome superconductors is the observation of the
Hebel-Slichter coherent peak in the temperature ($T$) dependence of
$(T_{1}T)^{-1}$. In a conventional $s$-wave superconductor, as shown
in Fig.~\ref{fig3}(a) for the result with $J=0.1$ which leads to the
conventional $s$-wave pairing, the $T$ dependence of $(T_{1}T)^{-1}$
develops a peak structure below $T_{c}$, which can be explained
theoretically as a result of the nonzero coherent factor described
in BCS theory along with the enhancement of the SC DOS at the gap
edge. Derivations of $(T_{1}T)^{-1}$ for the Hamiltonian (7) can be
found in the Appendix A. In the frustrated SC state, although the
amplitude of SC pairing modulates on the Fermi surface and the phase
of the SC pairing changes $4\pi$ around the Fermi surface, the phase
is the same at each pair of opposite van Hove singularities as
displayed in Fig.~\ref{fig2}(b), guaranteed by the sublattice
interference effect. This in turn leads to a nonzero coherent factor
in $(T_{1}T)^{-1}$ even within canonical SC states exhibiting
unconventional pairing symmetries~\cite{Jiang1,YDai1}. As expected,
the result with $J=0.2$ shown in Fig.~\ref{fig3}(b) clearly shows
the pronounced Hebel-Slichter peak of $(T_{1}T)^{-1}$ just below
$T_{c}$. The presence of the Hebel-Slichter peak in the frustrated
SC state keeps in line with the corresponding experiments.

In kagome superconductors, another important piece of evidence in
favor of conventional SC pairing is the absence of in-gap states
near a nonmagnetic impurity. In the presence of impurities, the
Hamiltonian should be varied by adding a impurity-scattering term
$H_{imp}=\sum_{i,\sigma}U_{i}c^{\dag}_{i\sigma}c_{i\sigma}$ with
$U_{i}$ denoting the scattering strength of the impurity on site
$i$.  In this circumstance, we should transform the Bogoliubov-de
Gennes equations in Eq. (11) into the real space as,
\begin{eqnarray}
\sum_{j}\left(
\begin{array}{cc}
H_{ij,\sigma} &
\Delta_{ij} \\
\Delta^{\ast}_{ij} & -H^{\ast}_{ij,\bar{\sigma}}
\end{array}
\right)\left(
\begin{array}{cc}
u_{n,j,\sigma} \\
v_{n,j,\bar{\sigma}}
\end{array}
\right)= E_{n}\left(
\begin{array}{cc}
u_{n,i,\sigma} \\
v_{n,i,\bar{\sigma}}
\end{array}
\right),
\end{eqnarray}
where
$H_{ij,\sigma}=-t_{ij}\delta_{i+\mathbf{\tau}_{j},j}-\mu\delta_{i,j}+U_{i}\delta_{i,j}$
with $\mathbf{\tau}_{j}$ denoting the four NN vectors.
$\Delta_{ij}=(\Delta_{i}-\frac{J}{V_{s}}\eta_{i})\delta_{i,j}$
represents the contributions from the on-site SC pairing and pairing
hoping. $u_{n,i,\sigma}$ and $v_{n,i,\bar{\sigma}}$ are the
Bogoliubov quasiparticle amplitudes on the $i$-th site with
corresponding eigenvalue $E_{n}$. The self-consistent equations for
the amplitude of the SC pairing $\Delta_{i}$ and the electron
density $n_{i}$ on site $i$ can be obtained in the same way as Eq.
(14). Then, the local density of state (LDOS) is given by
$\rho(E,\mathbf{r}_{i})=-\sum_{n}[|u_{n,i,\uparrow}|^{2}f^{'}(E_{n}-E)
+|v_{n,i,\downarrow}|^{2} f^{'}(E_{n}+E)]$, which is proportional to
the differential tunneling conductance observed in scanning
tunneling microscopy (STM) experiments. To diminish the finite size
effect, we adopt the supercell technique for the numerical
calculations in real space.

We consider the case of one unitary nonmagnetic impurity embodied in
the system, which is represented by setting
$U_{i}=U_{0}\delta_{i,i_{0}}$ with $U_{0}$ and $i_{0}$ denoting the
scattering strength and the location of the impurity. For
definition, we choose $U_{0}=50.0$ that is well in the unitary
regime and lay the impurity on the $A$ sublattice site as marked by
$A_{0}$ in Fig.~\ref{fig1}(a). Figs.~\ref{fig3}(c) and (d) present
respectively the LDOS results for $J=0.1$ and $J=0.2$ in the case of
one nonmagnetic impurity.

On the one hand, the bulk density of states (DOS) for $J=0.1$
exhibits a U-shaped profile characteristic of isotropic $s$-wave SC
pairing, while the basinlike V-shaped DOS for $J=0.2$ indicates
anisotropic yet fully gapped SC pairing, both behaviors aligning
with the expectations from Figs.~\ref{fig1}(b) and ~\ref{fig2}(a).
Despite these distinct pairing symmetries, neither the isotropic
$s$-wave nor the anisotropic frustrated SC state shows significant
impurity-induced in-gap states, as evidenced by the absence of such
features in Figs.~\ref{fig3}(c) and (d). The absence of the
impurity-induced in-gap states in the self-consistent solutions
corroborates the prediction that all conceivable unconventional
spin-singlet pairings in the kagome superconductors would produce a
disorder-response qualitatively similar to the conventional $s$-wave
superconductors~\cite{Holb1}.

On the other hand, while impurity-induced suppression of the SC gap
edges in the LDOS persists over long distances at the same
sublattice as the impurity site, it has minimal impact on LDOS at
different sublattice sites, even within the same unit cell. This
contrast underscores the distinct sublattice interference effect
inherent to the kagome system. Owing to sublattice interference,
each of the three van Hove singular points ($M_{A},M_{B},M_{C}$)
originates exclusively from one of the three inequivalent lattice
sites ($A,B,C$). Consequently, an impurity located at a specific
sublattice site, e.g., sublattice $A$, primarily scatters electrons
near the $M_{A}$ points in reciprocal space, with a corresponding
scattering effect observed in real space (See Appendix B for
details.). Thus, the impurity exerts a negligible influence on the
LDOS across different sublattices. Meanwhile, since the same
sublattice sites share identical SC phases, impurity effects
primarily manifest as suppression of the SC gap edges at the same
sublattice sites without inducing significant shifts of these edges
(With the exception of the same sublattice site within the NN unit
cell, which exhibits the strongest impurity-induced response.). This
behavior is quantitatively captured by curves $A_{2}-A_{5}$ in
Figs.~\ref{fig3}(c) and (d).

Theoretically, the local electronic structure at the impurity's
NN-site, which usually yield the optimal outcome, are widely
utilized to characterize such perturbations. However, in kagome
systems, this paradigm requires revision, since the strongest
impurity-induced signatures emerge not at the impurity's NN site,
but at the same sublattice of the NN unit cell, as unambiguously
demonstrated by the $A_{1}$ curves in Figs.~\ref{fig3}(c) and (d).

\vspace*{.0cm}

\subsection{Impurity-induced phase transition}
\begin{figure}[htb]
\begin{center}
\vspace{-.2cm}
\includegraphics[width=240pt,height=160pt]{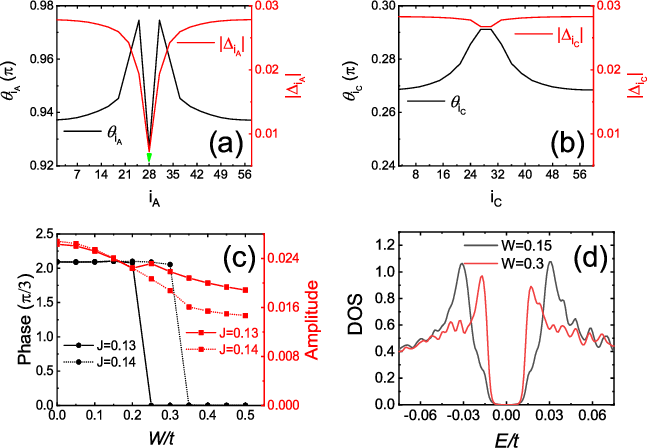}
\caption{Spatial distributions of SC amplitude $|\Delta_{i}|$ and
phase $\theta_{i}$ on the $A$-sublattice sites (a), and on the
$B$-sublattice sites (b), where a $U_{0}=2$ impurity is embedded at
the $A$-sublattice site indicated by the short green arrow in (a).
(c) The average phase difference $\varphi$ between the sublattices
and the average amplitude of the SC pairings $|\bar{\Delta}|$ as a
function of the impurity strength $W$. (d) The energy dependence of
the average DOS at different $W$ for $J=0.13$.}\label{fig4}
\end{center}
\end{figure}

Although the frustrated SC state remains insensitive to a strong
impurity manifested by the absence of impurity-induced in-gap
states, even a weak impurity does depress the SC amplitude
$|\Delta_{i}|$ and disturb the SC phase $\theta_{i}$ significantly
at the same sublattice sites near the impurity, as revealed in
Fig.~\ref{fig4}(a). Whereas the SC amplitude $|\Delta_{i}|$ exhibits
minor spatial variations across different sublattice sites near the
impurity, the SC phase $\theta_{i}$ also demonstrates a pronounced
deviation from its equilibrium value in the geometrically frustrated
SC state, as shown in Fig.~\ref{fig4}(b). These perturbations will
locally disrupt the delicate balance among the three adjacent
sublattices in the frustrated SC state. Thereby, the collective
effects arising from multiple impurities are expected to profoundly
modify the global configuration of the frustrated SC state.

To explicitly investigate this phenomenon, we consider the case of
random impurities which is defined by an independent random variable
$U_{i}$ randomly distributed over $[-W,W]$ at randomly selected half
sites $r_{i}$ of the total sites. Then $W$ is a parameter for
characterizing the strength of the disorder. For illustration
purposes, we define the average phase difference between the
sublattices as
$\varphi=\frac{1}{4N}\Sigma_{i}[(\theta_{A,i}-\theta_{B,i})+(\theta_{B,i}-\theta_{C,i})
+(\theta_{C,i}-\theta_{A,i})]$, and the average amplitude of the SC
pairings as
$|\bar{\Delta}|=\frac{1}{3N}\Sigma_{\alpha,i}|\Delta_{\alpha,i}|$,
where $\theta_{\alpha,i}$ and $N$ stand for respectively the phase
of the SC pairing on sublattice $(\alpha,i)$ and the number of the
total unit cells used in the calculations. Obviously, $\varphi$
should be equal or approach to $\frac{2\pi}{3}$ if the system is in
the frustrated SC state, and should be zero if it is in the
conventional $s$-wave state. Fig.~\ref{fig4}(c) presents the
disorder strength $W$ dependence of the phase difference $\varphi$
and the average amplitude $\bar{\Delta}$ for specific values of
$J=0.13$ and $J=0.14$. As clearly shown in the figure, while
$|\bar{\Delta}|$ exhibits a steady but weak decrease with disorder,
the phase difference $\varphi$ switches from $2\pi/3$ to zero at
$W=0.25$ for $J=0.13$ and at $W=0.35$ for $J=0.14$. Accordingly, as
displayed in Fig.~\ref{fig4}(d), the DOS undergoes a transition from
the basinlike V-shaped DOS at weak disorder to the U-shaped one at
strong disorder, portraying an impurity-induced SC pairing
transition from the anisotropic time-reversal symmetry breaking SC
state to the conventional $s$-wave SC state without passing through
the nodal point. It is worth noting that the disorder-induced SC
pairing transition mirrors the quintessential feature of the
experimental observation~\cite{Roppongi1}.

\vspace*{.0cm}

\subsection{Discussion}

The AV$_{3}$Sb$_{5}$ superconductors, which are kagome lattices near
the van Hove filling, would be the promising system for realizing
the frustrated superconductivity outline here. In these systems, the
sublattice interference near the van Hove singularity has a
pronounced effect on reducing the on-site interaction relative to
nearest neighbor interactions~\cite{Kiesel3}. This subsequently
facility the nearest neighbor Cooper pair hoppings in the
superconductors, which is precisely the requisite condition for the
frustrated superconductivity identified here.

The pair hopping interaction on the square lattice has been proposed
earlier to account for the $\eta$-type pairing phase where the total
momentum of the paired electrons is $Q=(\pi,\pi)$ and the phase of
SC order parameter alters from one site to the neighboring
one~\cite{Penson1,Bulka1,Czart1,Czart2,Japar1}. It was shown that
flux quantization and Meissner effect appear in this
state~\cite{Yang1,Yang2}. On the contrary, the alterations of the SC
phase between the three sublattices for the kagome lattice do not
break the translational symmetry, so the momentum of the SC pairing
still remains zero, just like the traditional SC state.
Nevertheless, the alteration of the SC phase causes the frustrated
superconductivity which simultaneously breaks time-reversal
symmetry. Another perspective posits that the phase misalignment of
SC pairings across distinct bands in multiband systems may induce a
time-reversal symmetry-broken SC state~\cite{Yerin1,Yerin2}, where
phase frustration originates from interband interaction effects
under specific conditions. By contrast, the frustrated SC state
proposed here arises from lattice topology constraints, i.e., the
geometric frustration effect. This fundamental distinction
highlights a novel mechanism where electronic phase coherence
competes with unique lattice geometry.

In a specific case, if we assign the phases $e^{-i2\pi/3}$, $1$ and
$e^{i2\pi/3}$ to the SC order parameters on the three sublattice
sites $A$, $B$ and $C$, respectively. In sublattice space, the SC
pairings can be written in the following matrix form,
\begin{eqnarray}
\left(
\begin{array}{ccc}
\Delta e^{-i\frac{2\pi}{3}} & 0 & 0 \\
0 & \Delta & 0 \\
0 & 0 & \Delta e^{i\frac{2\pi}{3}}
\end{array}
\right),
\end{eqnarray}
which is in turn decoupled into the real and imaginary parts as,
\begin{eqnarray}
\frac{1}{2}\left(
\begin{array}{ccc}
-\Delta & 0 & 0 \\
0 & 2\Delta & 0 \\
0 & 0 & -\Delta
\end{array}
\right)+i\frac{\sqrt{3}}{2}\left(
\begin{array}{ccc}
-\Delta & 0 & 0 \\
0 & 0 & 0 \\
0 & 0 & \Delta
\end{array}
\right).
\end{eqnarray}
This equivalents to the pairing symmetry of
$E^{(1)}_{2}+iE^{(2)}_{2}$ used in Ref.~\onlinecite{Holb1} up to a
global phase factor $e^{i\pi}$.

\section{Conclusion}

In conclusion, we have proposed a theory to explore the possible
frustrated SC state in the kagome superconductors. It is revealed in
this theory that the frustrated SC state emerges as a result of the
symmetry mismatch between the conventional $s$-wave SC pairing and
the crystal lattice when the NN-sites Cooper pairing hoping was
considered. In this state, the SC pairing on the three sublattices
has $2\pi/3$ phase difference from each other, leading to the
six-fold modulation of the amplitude and the time-reversal symmetry
breaking with $4\pi$ phase changes of the SC pairing as one
following it around the Fermi surface. In this state, the SC pairing
on the three sublattices exhibits a mutual $2\pi/3$ phase
difference. This phase arrangement leads to a six-fold modulation of
the SC order parameter amplitude and induces time-reversal symmetry
breaking through a $4\pi$ phase winding of the SC pairing when
traversing a closed path along the Fermi surface. On the other hand,
the frustrated SC state manifests a pronounced Hebel-Slichter
coherence peak in the nuclear spin-lattice relaxation rate below
$T_{c}$, and avoids impurity-induced in-gap states.In particular,
the theory also revealed a disorder-induced SC pairing transition
from the frustrated SC state to an isotropic $s$-wave SC state
without passing through the nodal point, a process that aligns with
and explains experimental observations. This theory not only offered
a promising framework to reconcile divergent or seemingly
contradictory experimental observations regarding SC properties, but
might also set the foundation for investigating geometrically
frustrated SC states in kagome lattice materials.

\section{acknowledgement}
\par
This work was supported by the NSFC/RGC JRS grant (No. N HKU774/21),
the GRF of Hong Kong (No. 17303023), the National Key Projects for
Research and Development of China (No. 2024YFA1408104) and the
National Natural Science Foundation of China (No. 12374137, and No.
12434005).


\appendix


\renewcommand{\thefigure}{\Alph{section}\arabic{figure}}

\hspace{1.6cm}

\section{Derivation of the spin-lattice relaxation rate}

The spin-lattice relaxation rate $\frac{1}{T_{1}T}$ is related to
the imaginary part of spin susceptibility,
\begin{eqnarray}
\frac{1}{T_{1}T}&\propto &\lim_{\omega\rightarrow
0}\frac{1}{N}\sum_{\mathbf{q}}\textmd{Im}\frac{\chi_{0}^{+-}(\omega,\mathbf{q})}{\omega}.
\end{eqnarray}
Here, $\chi_{0}^{+-}(\omega,\mathbf{q})=\int_{0}^{\beta}d\tau
e^{i\omega\tau}\chi_{0}^{+-}(\tau,\mathbf{q})$ with
$\chi_{0}^{+-}(\tau,\mathbf{q})$ defined as,
\begin{eqnarray}
\chi_{0}^{+-}(\tau,\mathbf{q})&=&\frac{1}{N}\langle
T_{\tau}S_{\mathbf{q}}^{+}(\tau)S_{-\mathbf{q}}^{-}(0)\rangle,
\end{eqnarray}
where
$S_{\mathbf{q}}^{+}=\Sigma_{l,\mathbf{k}}c_{l,\mathbf{k}+\mathbf{q},\uparrow}^{\dagger}
c_{l,\mathbf{k},\downarrow}$ and
$S_{\mathbf{-q}}^{-}=(S_{\mathbf{q}}^{+})^{\dagger}$. Resorting to
the Wick's theorem and making the Bogoliubov transformations
$c_{l,\mathbf{k},\uparrow}=\sum_{n}[u_{ln}(\mathbf{k})\gamma_{n\uparrow}(\mathbf{k})
-v_{ln}^{\ast}(\mathbf{k})\gamma_{n\downarrow}^{\dagger}(-\mathbf{k})]$
and
$c_{l,-\mathbf{k},\downarrow}=\sum_{n}[u_{ln}(\mathbf{k})\gamma_{n\downarrow}(-\mathbf{k})
+v_{ln}^{\ast}(\mathbf{k})\gamma_{n\uparrow}^{\dagger}(\mathbf{k})]$,
we finally arrive at the following expression of
$\chi_{0}^{+-}(\omega,\mathbf{q})$,
\begin{widetext}
\begin{eqnarray}
\chi_{0}^{+-}(\omega,\mathbf{q})=&&\frac{1}{N}\sum_{l\lambda,nm,\mathbf{k}}
[(u_{l,n,\mathbf{k}+\mathbf{q}}^{\ast}u_{\lambda,
n,\mathbf{k}+\mathbf{q}}v_{l,m,\mathbf{k}}^{\ast}v_{\lambda,
m,\mathbf{k}}-u_{l,n,\mathbf{k}+\mathbf{q}}^{\ast}v_{\lambda,
n,\mathbf{k}+\mathbf{q}}v_{l,m,\mathbf{k}}^{\ast}u_{\lambda,m,\mathbf{k}})
\frac{1-f(E_{n,\mathbf{k}+\mathbf{q}})-f(E_{m,\mathbf{k}})}
{\omega+E_{n,\mathbf{k}+\mathbf{q}}+E_{m,\mathbf{k}}+i\delta} \nonumber\\
&&+(v_{l,n,\mathbf{k}+\mathbf{q}}v_{\lambda,
n,\mathbf{k}+\mathbf{q}}^{\ast}v_{l,m,\mathbf{k}}^{\ast}v_{\lambda,
m,\mathbf{k}}+v_{l,n,\mathbf{k}+\mathbf{q}}u_{\lambda,
n,\mathbf{k}+\mathbf{q}}^{\ast}v_{l,m,\mathbf{k}}^{\ast}u_{\lambda,m,\mathbf{k}})
\frac{f(E_{n,\mathbf{k}+\mathbf{q}})-f(E_{m,\mathbf{k}})}
{\omega-E_{n,\mathbf{k}+\mathbf{q}}+E_{m,\mathbf{k}}+i\delta} \nonumber\\
&&+(u_{l,n,\mathbf{k}+\mathbf{q}}^{\ast}u_{\lambda,
n,\mathbf{k}+\mathbf{q}}u_{l,m,\mathbf{k}}u_{\lambda,
m,\mathbf{k}}+u_{l,n,\mathbf{k}+\mathbf{q}}^{\ast}v_{\lambda,
n,\mathbf{k}+\mathbf{q}}u_{l,m,\mathbf{k}}v_{\lambda,m,\mathbf{k}}^{\ast})
\frac{f(E_{m,\mathbf{k}})-f(E_{n,\mathbf{k}+\mathbf{q}})}
{\omega-E_{m,\mathbf{k}}+E_{n,\mathbf{k}+\mathbf{q}}+i\delta} \nonumber\\
&&+(v_{l,n,\mathbf{k}+\mathbf{q}}v_{\lambda,
n,\mathbf{k}+\mathbf{q}}^{\ast}u_{l,m,\mathbf{k}}u_{\lambda,
m,\mathbf{k}}^{\ast}-v_{l,n,\mathbf{k}+\mathbf{q}}u_{\lambda,
n,\mathbf{k}+\mathbf{q}}^{\ast}u_{l,m,\mathbf{k}}v_{\lambda,m,\mathbf{k}}^{\ast})
\frac{f(E_{n,\mathbf{k}+\mathbf{q}})+f(E_{m,\mathbf{k}})-1}
{\omega-E_{n,\mathbf{k}+\mathbf{q}}-E_{m,\mathbf{k}}+i\delta}].
\end{eqnarray}
\end{widetext}

\section{Sublattice interference effect on the impurity state}

The observed sublattice dichotomy, where impurity-induced states
strongly localize on the host sublattice while leaving the LDOS on
the adjacent sublattice nearly intact, is qualitatively explained by
the $T$-matrix formalism through sublattice interference processes.

Within the $T$-matrix framework, the impurity-renormalized Green's
function at site $\mathbf{r}$ is formally expressed as,
\begin{widetext}
\begin{eqnarray}
\hat{G}(\mathbf{r},\mathbf{r},i\omega_{n})&=&\hat{G}_{0}(\mathbf{r}-\mathbf{r},i\omega)
+\hat{G}_{0}(\mathbf{r}-\mathbf{r}_{0},i\omega)\hat{T}(\mathbf{r}_{0},i\omega)
\hat{G}_{0}(\mathbf{r}_{0}-\mathbf{r},i\omega),
\end{eqnarray}
\end{widetext}
where $\hat{G}_{0}(i\omega)$ denotes the bare Green's function, and
the $\hat{T}(\mathbf{r}_{0},i\omega)$-matrix encapsulates the full
scattering sequence through the Dyson series,
\begin{eqnarray}
\hat{T}(\mathbf{r}_{0},i\omega)&=&\frac{\hat{V}}{\hat{1}-
\hat{V}\hat{G}_{0}(\mathbf{r}_{0},i\omega)}.
\end{eqnarray}
Here, $\hat{V}=V_{0}\hat{\tau}_{3}$ accounts for the potential
impurity at location $\mathbf{r}_{0}$. Then, the LDOS at site
$\mathbf{r}$ is expressed as
$\rho(E,\mathbf{r})=-\frac{1}{\pi}\textmd{Im}
[\hat{G}_{ee}(\mathbf{r},\mathbf{r},\omega_{n}+i\delta)
+\hat{G}_{hh}(\mathbf{r},\mathbf{r},-\omega_{n}+i\delta)]$, where
the diagonal elements $\hat{G}_{ee}$ and $\hat{G}_{hh}$ correspond
to the contributions from the electron and hole parts, respectively.

\setcounter{figure}{0}
\begin{figure}[h]
\begin{center}
\vspace{.6cm}
{\mbox{\includegraphics[width=230pt,height=160pt]{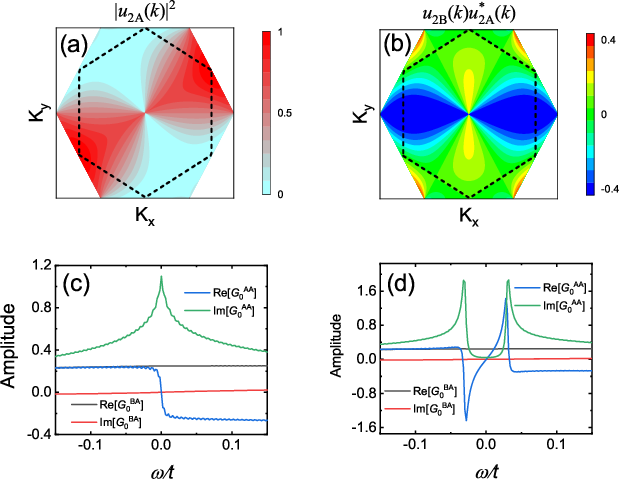}}}
\caption{The momentum distribution of the form factors
$|u_{2,A}(\mathbf{k})|^{2}$ (a), and
$u_{2,B}(\mathbf{k})u_{2,A}^{\ast}(\mathbf{k})$ (b), respectively.
Real and imaginary parts of the Green's functions
$\hat{G}_{0,ee}^{AA}(\mathbf{r}_{A_{1}}-\mathbf{r}_{A_{0}},\omega+i\delta)$
and
$\hat{G}_{0,ee}^{BA}(\mathbf{r}_{B_{0}}-\mathbf{r}_{A_{0}},\omega+i\delta)$
for the normal state (c), and for the SC state (d).
$\hat{G}_{0,ee}^{AA}(\mathbf{r}_{A_{1}}-\mathbf{r}_{A_{0}},\omega+i\delta)$
and
$\hat{G}_{0,ee}^{BA}(\mathbf{r}_{B_{0}}-\mathbf{r}_{A_{0}},\omega+i\delta)$
have been abbreviated respectively as $\hat{G}_{0}^{AA}$ and
$\hat{G}_{0}^{BA}$ in the figures.} \label{fig5}
\end{center}
\end{figure}

For an impurity with a given strength located on a specific
sublattice site $A$ at location $\mathbf{r}_{A_{0}}$, the Green's
functions at the same sublattice $A$ at location $\mathbf{r}_{A}$ is
expressed as
\begin{widetext}
\begin{eqnarray}
\hat{G}^{AA}(\mathbf{r}_{A},\mathbf{r}_{A},i\omega_{n})&=&\hat{G}_{0}^{AA}(\mathbf{r}_{A}-\mathbf{r}_{A},i\omega)
+\hat{G}_{0}^{AA}(\mathbf{r}_{A}-\mathbf{r}_{A_{0}},i\omega)\hat{T}(\mathbf{r}_{A_{0}},i\omega)
\hat{G}_{0}^{AA}(\mathbf{r}_{A_{0}}-\mathbf{r}_{A},i\omega),
\end{eqnarray}
\end{widetext}
while the Green's functions at another sublattice $B$ at location
$\mathbf{r}_{B}$ is expressed as
\begin{widetext}
\begin{eqnarray}
\hat{G}^{BB}(\mathbf{r}_{B},\mathbf{r}_{B},i\omega_{n})&=&\hat{G}_{0}^{BB}(\mathbf{r}_{B}-\mathbf{r}_{B},i\omega)
+\hat{G}_{0}^{BA}(\mathbf{r}_{B}-\mathbf{r}_{A_{0}},i\omega)\hat{T}(\mathbf{r}_{A_{0}},i\omega)
\hat{G}_{0}^{AB}(\mathbf{r}_{A_{0}}-\mathbf{r}_{B},i\omega).
\end{eqnarray}
\end{widetext}
Both the bare Green's functions
$\hat{G}_{0}^{AA/BB}(\mathbf{r}_{A/B}-\mathbf{r}_{A/B},i\omega)$ and
the $\hat{T}(\mathbf{r}_{0}=\mathbf{r}_{A_{0}},i\omega)$-matrix in
Eqs. (A3) and (A4) are the same, so the difference comes from
$\hat{G}_{0}^{AA}(\mathbf{r}_{A}-\mathbf{r}_{A_{0}},i\omega)$ and
$\hat{G}_{0}^{BA}(\mathbf{r}_{B}-\mathbf{r}_{A_{0}},i\omega)$. For
simplicity, we only consider the electron part of
$\hat{G}_{0}(i\omega)$ in the following.
\begin{eqnarray}
\hat{G}_{0,ee}^{l\lambda}(\mathbf{r}-\mathbf{r}_{0},i\omega)=-\int_{0}^{\beta}e^{i\omega\tau}
d\tau\langle
T_{\tau}c_{l,\mathbf{r}}^{\dag}(\tau)c_{\lambda,\mathbf{r}_{0}}(0)\rangle.
\end{eqnarray}
After making the transformations
$c_{\lambda,\mathbf{r}}=\sum_{k}c_{\lambda,\mathbf{k}}e^{i\mathbf{k}\cdot\mathbf{r}}$
and
$c_{\lambda,\mathbf{k}}=\sum_{n}u_{n,\lambda}^{\ast}(\mathbf{k})\phi_{n}(\mathbf{k})$,
we finally arrive at,
\begin{eqnarray}
\hat{G}_{0,ee}^{AA}(\mathbf{r}_{A}-\mathbf{r}_{A_{0}},i\omega)=
\sum_{n,\mathbf{k}}\frac{|u_{n,A}(\mathbf{k})|^{2}}
{i\omega-\varepsilon_{n}(\mathbf{k})}e^{-i\mathbf{k}\cdot(\mathbf{r}_{A}-\mathbf{r}_{A_{0}})},
\end{eqnarray}
and
\begin{eqnarray}
\hat{G}_{0,ee}^{BA}(\mathbf{r}_{B}-\mathbf{r}_{A_{0}},i\omega)=
\sum_{n,\mathbf{k}}\frac{u_{n,B}(\mathbf{k})u_{n,A}^{\ast}(\mathbf{k})}
{i\omega-\varepsilon_{n}(\mathbf{k})}e^{-i\mathbf{k}\cdot(\mathbf{r}_{B}-\mathbf{r}_{A_{0}})}.
\end{eqnarray}
Since the band $2$ crosses the Fermi energy, the form factors
$|u_{2,A}(\mathbf{k})|^{2}$ and
$u_{2,B}(\mathbf{k})u_{2,A}^{\ast}(\mathbf{k})$ are dominant in Eqs.
(B6) and (B7). Figs.~\ref{fig5}(a) and (b) presents the momentum
distribution of the form factors $|u_{2,A}(\mathbf{k})|^{2}$ and
$u_{2,B}(\mathbf{k})u_{2,A}^{\ast}(\mathbf{k})$.

Owing to the sublattice interference effect, the form factor
$|u_{2,A}(\mathbf{k})|^{2}$ peaks predominantly around van Hove
singularities, reaching maximal intensity precisely at these
critical points. The overlapping of the two factors will boost the
Green's function
$\hat{G}_{0,ee}^{AA}(\mathbf{r}_{A}-\mathbf{r}_{A_{0}},\omega+i\delta)$
to a significant changes near the Fermi energy for the normal state
and near the gap edge for the SC state, as shown in
Fig.~\ref{fig5}(c) and (d) for the Green's functions between the
impurity and the same sublattice of the NN-unit cell
$\hat{G}_{0,ee}^{AA}(\mathbf{r}_{A_{1}}-\mathbf{r}_{A_{0}},\omega+i\delta)$.
On the contrary, the form factor
$u_{2,B}(\mathbf{k})u_{2,A}^{\ast}(\mathbf{k})$ is zero along most
part of the Fermi surface except for the segments crossing the blue
regions, exactly avoiding the van Hove singularities. The
coincidence of the nonzero form factor and low DOS segments of the
Fermi surface causes featureless for the Green's function
$\hat{G}_{0,ee}^{BA}(\mathbf{r}_{B}-\mathbf{r}_{A_{0}},i\omega)$, as
displayed in Fig.~\ref{fig5}(c) and (d) for the Green's functions
between the impurity and its NN $B$ sublattice site
$\hat{G}_{0,ee}^{BA}(\mathbf{r}_{B_{0}}-\mathbf{r}_{A_{0}},\omega+i\delta)$.
Therefor, the impurity produces the strongest impurity-induced
response at the same sublattice site in the NN unit cell, while it
exerts a negligible influence on the LDOS across different
sublattices.

\end{document}